\documentclass{svmult}

\usepackage{type1cm}        
\usepackage{makeidx}         
\usepackage{graphicx}        
\usepackage{multicol}        
\usepackage[bottom]{footmisc}

\usepackage{newtxtext}       %
\usepackage[varvw]{newtxmath}       

\usepackage[colorlinks=true,linkcolor=blue,urlcolor=blue,allcolors=blue]{hyperref}

\makeindex             

\begin{document}

\title*{Design of radiotelemetry systems for animal tracking}
\author{Laila D. Kazimierski\orcidID{0000-0002-1305-9293} \\Guillermo Abramson\orcidID{0000-0001-6208-966X}\\ Nicol\'{a}s Catalano}
\institute{L. D. Kazimierski\at Centro At\'{o}mico Bariloche and CONICET, Av. Bustillo 9500, R8402AGP Bariloche, Argentina, \email{kazimierski@cab.cnea.gov.ar}
\and G. Abramson \at Centro At\'{o}mico Bariloche, CONICET and Instituto Balseiro, Av. Bustillo 9500, R8402AGP Bariloche, Argentina,  \email{abramson@cab.cnea.gov.ar}
\and N. Catalano \at Centro At\'omico Bariloche - Comisi\'on Nacional de Energ\'{\i}a At\'omica (CNEA), Instituto Balseiro - Universidad Nacional de Cuyo - CNEA, R8402AGP Bariloche, Argentina \email{nicolas.catalano@cab.cnea.gov.ar}
}

%
%
\maketitle

\abstract*{We describe the design and implementation of two low-cost, low-weight, radiotelemetry systems to measure the movement of small animals in a dense forest, where satellite positioning systems are unreliable and the attenuation of the vegetation poses several challenges. Both methods use stationary receiving stations that record the signal emitted by a portable transmitter carried by the animal. One of the methods measures the received power, while the other one registers the phase difference received by two-antennas stations. The later overcomes several difficulties that exist in the determination of the distance by the power method. We used our system to record the movement of \emph{Dromiciops gliroides}, a vulnerable South American marsupial native of the Patagonian Andes, where it plays an important role in the ecosystem.}

\abstract{We describe the design and implementation of two low-cost, low-weight, radiotelemetry systems to measure the movement of small animals in a dense forest, where satellite positioning systems are unreliable and the attenuation of the vegetation poses several challenges. Both methods use stationary receiving stations that record the signal emitted by a portable transmitter carried by the animal. One of the methods measures the received power, while the other one registers the phase difference received by two-antennas stations. The later overcomes several difficulties that exist in the determination of the distance by the power method. We used our system to record the movement of \emph{Dromiciops gliroides}, a vulnerable South American marsupial native of the Patagonian Andes, where it plays an important role in the ecosystem.}

\newpage

\section{Introduction}
\label{sec:1}

The knowledge of animals' movement is a crucial step towards the understanding of their use of resources,  the role they play in their habitat and ecosystem, as well as the impact and challenges that human activities pose on them. Most animal species are able to perform complex patterns of movement that, in general, depend on the environment and their interactions with other animals, and on their internal state. In the case of foragers, for example, these trajectories depend strongly on the vegetation that provides the resources and constrain the movement of individuals. Some of these species play key ecological roles in biomes with impoverished faunas. Such is the case of the South American marsupial \emph{Dromiciops gliroides} \cite{rodriguezcabal2013,rodriguezcabal2007,fonturbel2012}, the only extant species of the order Microbiotheria (classified as ``near threatened'' by IUCN \cite{iucn} and ``vulnerable'' in Argentina \cite{diaz2000}). This small arboreal animal is endemic to the northern portion (around $41^{\circ}$ S) of the Patagonian temperate forest \cite{kelt1989,martin2010}, where \emph{D. gliroides} is considered an ``ecological architect'' \cite{fonturbel2011}. They are able to regulate the spatial distribution of at least 16 plant species, including the mistletoe \emph{Tristerix corymbosus} which, since it blossoms in winter, is a crucial source of resources for the hummingbird \emph{Sephanoides sephaniodes}, one of the main pollinators of the whole biome \cite{amico2000,amico2009,amico2011}. In this context, the knowledge of the basic biology of \emph{D. gliroides} is key to taking conservation measures, since both the felling of forests and the extraction of firewood, as well as the creation of new paths in the forest, could interfere with the marsupial's movements and its role as a seed disperser in its habitat \cite{oliver2017}. The disruption of the key interaction between \emph{D. gliroides} and \emph{T. corymbosus} may have cascading effects, leading to the dismantling of the entire ecological network, altering the various relationships between species in the community \cite{rodriguezcabal2013,aizen2003}.

\emph{Dromiciops gliroides} is nocturnal; their average weight is 30 grams and they live in a forest with dense vegetation. These characteristics present great challenges for the study of their movement and behavior in the field, for a variety of reasons. The density of vegetation makes it difficult to use satellite global positioning systems, like GPS, the equipment that each animal can carry must be extremely light, and measurements must be made at night when the animals are on the move. Previous attempts to monitor their movement have found difficulties in reducing the measurement errors, suggesting the need to continue improving the current monitoring systems. In Chile, radiotelemetry techniques were implemented to determine home ranges of \emph{D. gliroides} \cite{fonturbel2009,fonturbel2016}, and a recent study in Brazil has combined techniques for the study of the movement of a different forest mammal \cite{brigatti2024}. However there is still a need to develop appropriate monitoring systems, applicable to small animals, that are not extremely expensive \cite{kays2011}, and are adequate to operate in dense forests, where employing systems like GPS positioning is not reliable.

In Argentina, until the present work, the method of choice for the study of the movement, use and selection of habitat of \emph{D. gliroides}, has been the spool and line technique \cite{boonstra1986}, which allows recover the trajectory of the animal by following the thread through the vegetation. The coils of thread are fixed to the back of the animal by means of an adhesive and, prior to releasing the individuals, the tip of the thread is secured to the branch on which the release is performed at the beginning of the night. The thread is afterwards deployed by the animal while moving. This technique produces high resolution trajectories but has several limitations. For example, the equipment that the animal must carry weighs approximately 2.5 grams, which could affect their movements, since the maximum tolerable load for birds and mammals is equivalent to 5\% of their body weight \cite{white1990,millspaugh2001}. This also limits the length of the coil, which does not exceed 100 meters. Furthermore, this technique does not record temporal information regarding the movement: we cannot determine how long the animal stayed at each place nor how long the entire trajectory lasted.

Our work, summarized below, involved the design, development, characterization and implementation of a tracking system using radiotelemetry techniques for \emph{D. gliroides}, but adaptable to small animals that live in similar environments. These techniques provide solutions to the problems that arise from using the coils of thread: the equipment placed on the animal weighs just $0.65$ grams and the temporal information is recorded together with the spatial one. We developed two methodologies~ \cite{kazimierski2021,catalano2018}. The first one consists on measuring the power of a signal emitted by a transmitter attached to the animal, which depends on its distance from three fixed receiving stations. The second one complements this information with the determination of the angle of arrival of the signal. The data recorded in the forest is subsequently processed, determining the location of the animal and rebuilding its trajectory. In Section~\ref{sec:power} we explain the method based on the received power and the trilateration technique that determines the probable position for each pulse. In Section~\ref{sec:phase} we explain the method based on the phase difference of received signals, and the triangulation technique to determine the trajectory. The applicability of systems  with these characteristics is very broad, in particular in forested environments like the site of our field work, where there is still much to learn about the resident species.

\section{Power measurement}
\label{sec:power}
We show the technique used in the first stage of the experimental work, in which we measure the power of the signal emitted by a transmitter placed on the animal and received by three stationary radio stations. We detail the methodology, including results obtained at different stages of the system's calibration. 
The results of the field implementation of this methodology (and that explained in the following section) are presented together in section \ref{sec:results}.

\subsection{Equipment and methodology}

\begin{figure}[t]
\sidecaption[t]
\includegraphics[width=7cm]{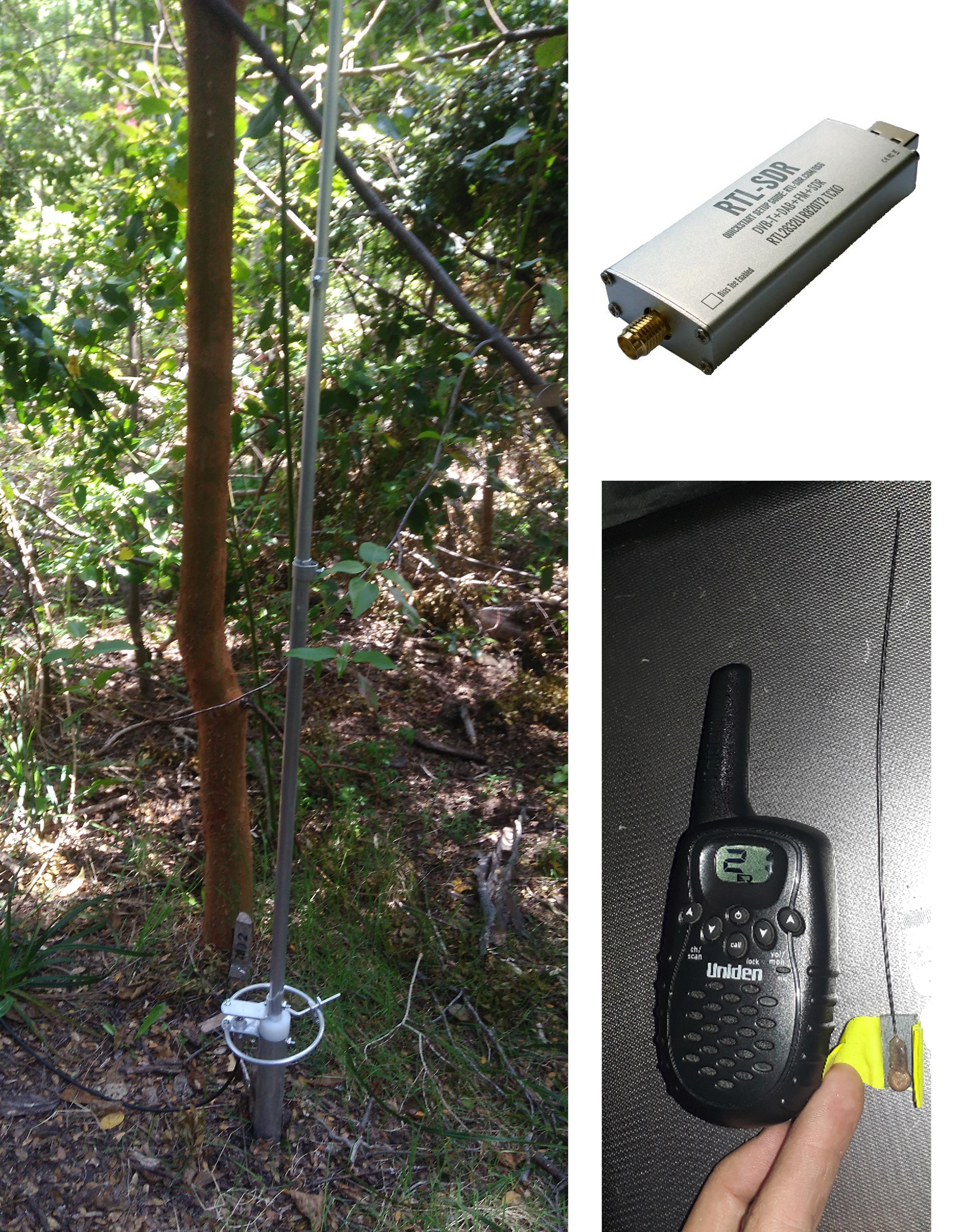}
\caption{Left: omnidirectional dipole antenna (Eiffel), fixed at ground level in the forest. That antenna is connected to a receiver which is connected to a laptop computer. Right, top: RTL-SDR dongle used to decode the received signal and feed it into the laptop computer. Right, bottom: one of the transmitters (ATS A2426) used on the animals, next to a walkie-talkie used to communicate in the field. These small transmitters emit periodic pulses every $4$ seconds at a frequency about $150$~MHz
, each of a duration of $18$~ms. They weigh $0.65$~g and the battery lasts $67$ days according to manufacturing data. Once they transmitters are activated, they continue pulsing until the battery runs out.}
\label{fig:antenna}       
\end{figure}

The system consists of an arrangement of three receiving stations, located on the edges of the area to be monitored. Each one has a commercial omnidirectional antenna, set to receive 150~MHz frequency signals emitted by the transmitter placed on the animal, an RTL-SDR receiver (Software Defined Radio) (see Figure~\ref{fig:antenna}) \cite{mishra2017,sdr}, and a laptop computer (that could eventually be replaced by smaller processing units). The antennas are connected to the receivers and these, in turn, to the processing units by means of a USB interface. The three stations simultaneously receive the pulses emitted by the transmitter placed on the animal.

Each individual whose movement is to be monitored has a transmitter weighing less than $1$~g, authorized for its use in species such as \emph{D. gliroides}, powered by a battery that allows it to emit pulses for a period of up to two months (depending on the battery model). To attach the transmitter, the fur is trimmed from the back of the animal, and the deviced is fixed with appropriate glue. After regrowth of the fur, the transmitter comes off on its own and is lost. We used Telenax (Querétaro, México), TXA-004G and ATS (Isanti, USA) A2426 transmitters, broadcasting at frequencies close to 150~MHz. These transmitters emit periodic pulses every 2 or 4 seconds depending on the brand and model, with a pulse duration of approximately 20~ms. (See Fig.~\ref{fig:antenna}.) 

\begin{figure}[t]
\centering
\includegraphics[width=\columnwidth]{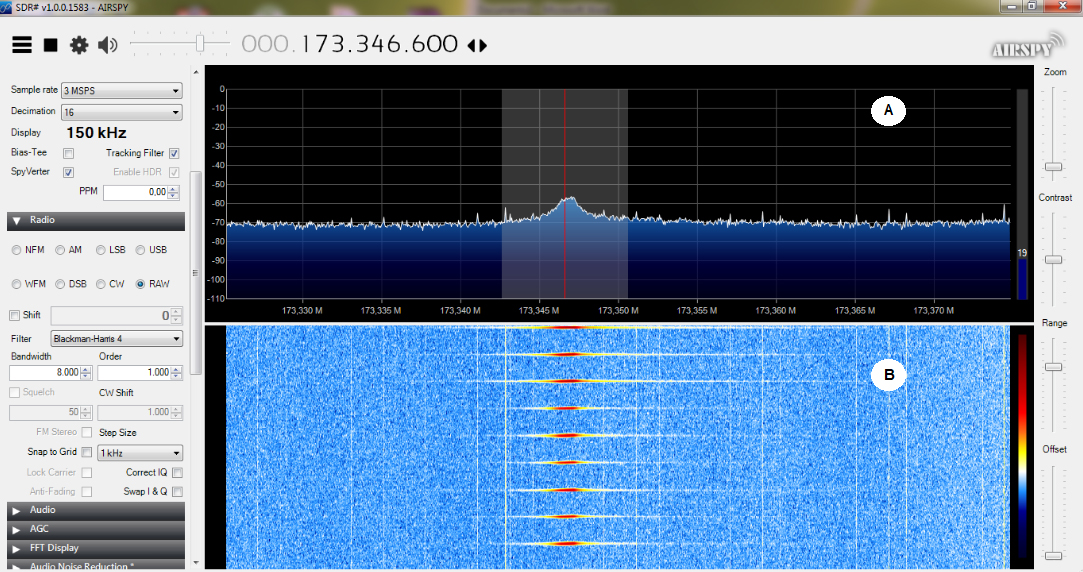}
\caption{Screen capture of the SDR\# during a test, emmitting pulses at 173 MHz. The left panel contains configuration options, such as the gain and recording modes. Panel A shows a pulse in real time, above background noise and within the selected frequency band width. Panel B shows a spectrogram of the last 40 s.}
\label{fig:sdr}       
\end{figure}

To check and record the pulses acquired by each station we used the software SDR\#; Figure \ref{fig:sdr} shows a capture during a test recording. The three receiving stations remain stationary during the measurement and between measurements, at positions that were precisely georeferenced with specialized equipment.

The emitted pulses are received by the three reception stations simultaneously with a certain power, which depends on the distance from the transmitter to each one. In an open space the power received by an antenna decays with the distance as~$r^{-2}$. However, in experiments carried out in the forest, we observed that this dependence is not strictly fulfilled. It is possible, nevertheless, to perform a calibration of the functional relationship of power with distance, which requires measuring the pulses emitted from at least two points located at known distances from the antennas~\cite{javaid2015,oguejiofor2013}.

\subsection{Forest trial}

Our work site was a one-hectare square portion of Patagonian temperate forest ($41 ^\circ\,02'\,57''$ S, $71^\circ\,32'\,44''$ W), previously georeferenced by a specialized surveyor, located in the Llao Llao Municipal Park, 30 km west of the city of Bariloche (Río Negro, Argentina). The site has a grid of $6\times 6$ marked pegs, separated 20 meters from each other, which we used to place the receiving stations. To capture suitable individuals to fit the transmitter on, we located 30 Tomahawk-type baited traps ($30\times 14\times 14$~cm) in the forest. The captured animals were fitted with a Telenax transmitter (similar to the ATS shown in Fig.~\ref{fig:antenna}) and released at a georeferenced point in the forest, to be measured during the hours of darkness, when they display greater activity.


Every night, before monitoring the animal, we performed a calibration of the system. The power of the emitted signal is affected by environmental conditions (temperature and humidity), as well as the charge of the battery, producing deviations from night to night, which need to be assessed. The calibration consists in associating the received power with the distance between the transmitter and each station. For this, we placed a transmitter (one that we used as reference, and let discharge in a way similar to the one placed on the animal) at three sites of the georeferenced grid and let it transmit pulses during one  minute, receiving them from the three stations. The recordings were used afterwards for the calibration of the signals transmitted by the animal. Since there is a dispersion in the power received during the minute of measurement, we had to establish some criterion to determine its value. Different possibilities include averaging all the pulses or choosing the maximum value, for example. The latter one is reasonable in our case, since the maximum is probably closest to the real one, given the attenuation produced by the high density of vegetation (assuming no constructive interference). Due to the heterogeneity of the environment, the function that best fits the relationship between power and distance depends on each receiving station. 


Once the calibration is complete, we begin to obtain measurements from the three receiving stations simultaneously during several hours. To coordinate the beginning of the measurements in all stations, and to align the pulses during the course of the measurement, we used a reference signal transmitted by a handheld transceiver (a ``walkie-talkie'', Fig.~\ref{fig:antenna}), at a different frequency but within the bandwidth captured by the receivers. The intensity of these pulses and their characteristic shape make them easy to differentiate from the transmitter's pulses.

\begin{figure}[t]
\includegraphics[width=\textwidth]{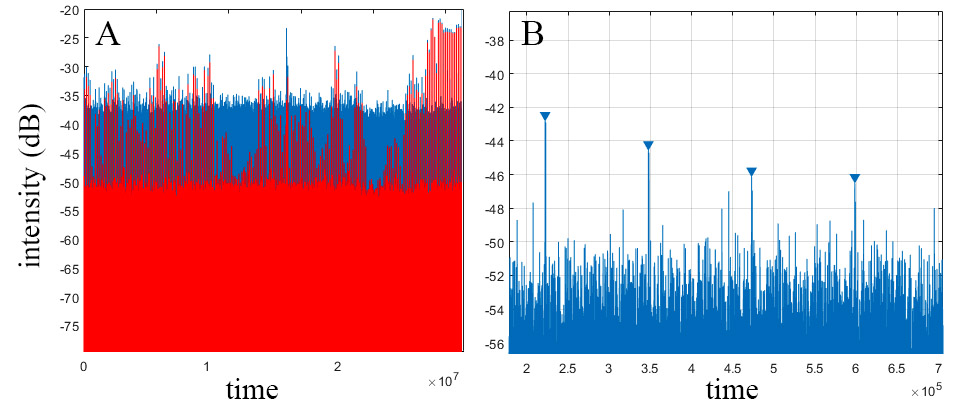}
\caption{A: signal measured by one of the stations (blue) and the same filtered using the low-pass filter (red). B: Pulse detection (marked by triangles) above background noise.}
\label{fig:signal}       
\end{figure}

We processed the recorded signals synchronizing the captures of the three stations, and filtering the noise using a low-pass filter, which is commonly used in signal filtering \cite{rani2017,mathworks}). By way of illustration, Figure \ref{fig:signal} shows the capture of one of the receiving stations corresponding to one night of measurement: in blue we show the original signal and, in red, the filtered signal. We observe that the noise of the original signal is significantly reduced. Once the signal is filtered, we detect the transmission pulses (see Fig.~\ref{fig:signal}, right) using different criteria. These peaks have a peculiar width and shape, which can be complemented with the use of a power threshold.

Since it can happen that some stations skip some pulses (because the transmitter is too far from them, among other possible reasons), we had to check for these events and discard the signal if at least one station has lost it. Only the simultaneous recording of three pulses can be used by the localization algorithm.

\begin{figure}[t]
\includegraphics[width=\textwidth]{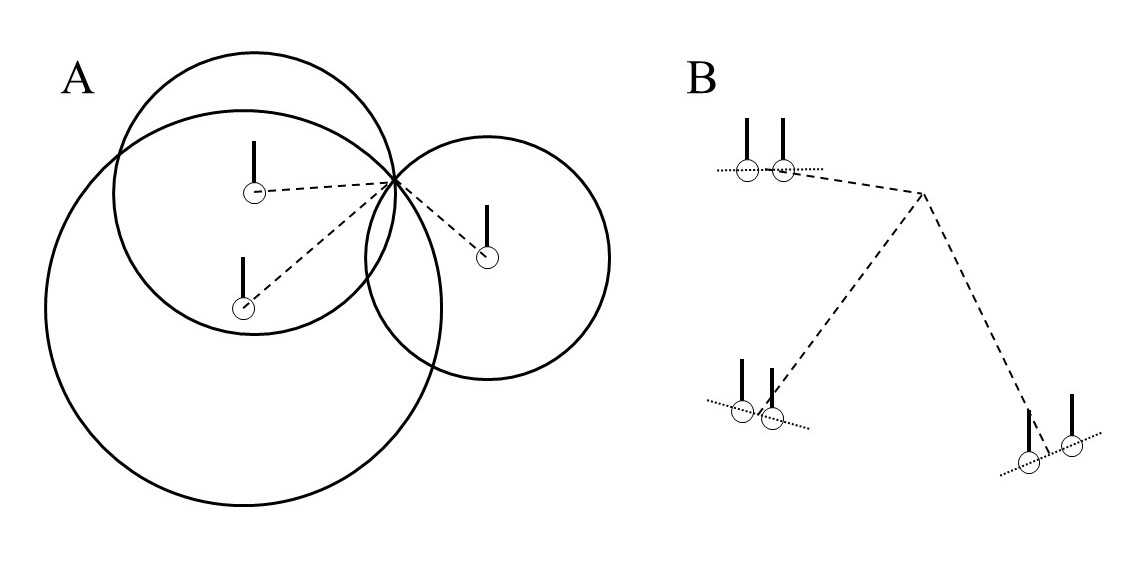}
\caption{A: Trilateration method, with the position of the transmitter identified as the intersection of three circles, centered at the three receiving antennas, with radius corresponding to the signal power. B: Triangulation method, with the transmitter located at the intersection of three lines, with their directions derived from the phase difference detected at each antenna of the three receiving station (described in Section \ref{sec:phase}).}
\label{fig:tri}       
\end{figure}

After these preliminaries, we can proceed with the determination of the animal position by trilateration, using the three circles shown in Fig.~\ref{fig:tri}. The most probable value is calculated using least squares, determining the point of intersection of the three circles or the closest point to them if there is no intersection. If we consider a temporal window in which we assume that the animal does not perform large displacements, we can use the maximum power value of each station within that range to trilateralize, rather than considering all points. The width of this window depends on how much the animal can move in a given time interval and the resolution that we can achieve with the methodology. The estimated average speed of \emph{D. gliroides} is reported at $5.79\pm 7.67$~m/min \cite{divirgilio2014}, allowing the use of 60~s temporal windows, with a resolution of 5 meters. The number of points to trilateralize, then, depends on the duration of the measurements and the temporal window. Using this criterion we could significantly reduce the dispersion of the data (relative to considering all pulses).
To overcome the limitations of the power measurement method, we have also explored an alternative based on the measurement of the phase difference, explained in the next section.

\section{Phase difference measurement}
\label{sec:phase}
With this methodology the signal emitted by the transmitter reaches two antennas located at the same station. This allows to calculate the direction of arrival of the signal and the reconstruction of the point of transmission. Below we detail the phase difference measurement.

\subsection{Equipment and methodology} 

\begin{figure}[t]
\includegraphics[width=\textwidth]{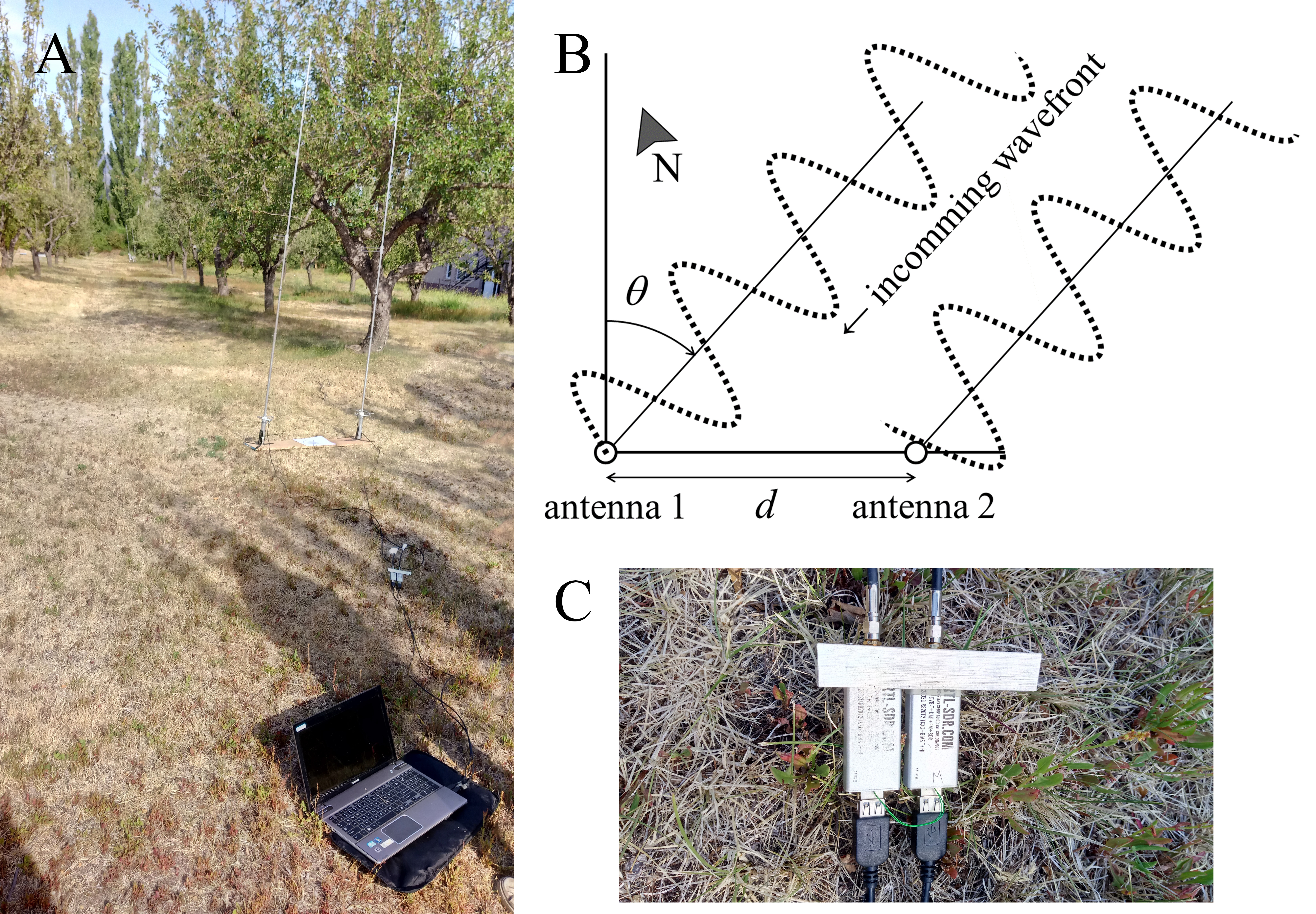}
\caption{Equipment  used for the phase difference measurement. A: Two antennas, placed on a rigid platform and connected to the same computer. B: Determination of the angle of arrival from the phase difference. C: Detail of the two RTL-SDR's sharing a single internal clock.}
\label{fig:2antennas}       
\end{figure}

The equipment layout is similar to that detailed in Section \ref{sec:power} but, in this case, each receiving station comprises two antennas connected to corresponding RTL-SDR receivers, which needed to be modified for this application, as discussed later. Figure \ref{fig:2antennas} shows the arrangement of a pair of antennas connected to a pair of receivers, in turn connected to a computer. In addition, we incorporated the use of low noise radiofrequency amplifiers (Qorvo MMIC SPF5189) to the equipment, which were connected between the antennas and the receivers.

The phase difference between the incoming signals arriving at the antennas can be derived from the wavelength and the separation between them, and allows to calculate the direction of arrival with respect to their baseline. This, in turn, can be converted to a direction of arrival with respect to the grid, given the azimuth of the baseline. 

The antennas were placed on a rigid platform to ensure a fixed distance between them. The variable to be determined with this configuration is the angle of arrival of the signal, $\theta$ (Figure \ref{fig:2antennas}), between the normal to the baseline and the incoming signal. If, in addition, we have the analogous information (i.e. position and angle of arrival) of the two remaining pairs of antennas, we can triangulate the position of the transmitter (Figure \ref{fig:tri} B). 

The path difference of the signal arriving at  each antenna is $d\sin\theta = \tau c$, where $\tau$ is the time delay, $c$ is the speed of light and $d$ is the separation between the antennas. This distance is set at half a wavelength of 150 Hz frequency signal we used, that is, $d = \lambda/2 = 1$~m. The time delay of the signal from one antenna to the other of the same pair implies a phase difference $\phi$ between the signals that reach the two antennas, equivalent to $\phi = 2\pi\tau/T$, where $T = \lambda/c$ is the period of the signal. Expressing $\phi$ as a function of $\theta$ we have:
\begin{align}
\phi = \frac{2\pi}{T}\frac{d\sin\theta}{c}=\pi\sin\theta.
\label{phase}
\end{align}

To measure the phase difference between the signals received by each pair of antennas, it is necessary to perform several synchronization steps: the two clocks of the receivers must be synchronized, the USB packets that the receivers send to the PC must also be synchronized, and the phase of the receiver must be known. Below, we detail each one of these.

First, to ensure that the sampling of the signals is carried out simultaneously and at the same frequency, we use the same clock for both receivers. To achieve this, we modified one of the receivers to use the sampling clock taken from the other receiver, as shown in Figure \ref{fig:2antennas}.

When employing non real time operating systems, as Microsoft Windows, sending a command to a USB device and receiving its answer involves a nondeterministic time. If we need that data coming from two devices to be time synchronized, we need to implement some way to achieve this synchronization manually as Windows doesn't provide means to do this.

Moreover, USB data packets do not have embedded temporal information, so it is not possible to align temporarily the received data coming from different devices. To cope with these synchronization problems, we sent an additional synchronization signal to both antennas so that, during the processing stage, we were be able to align the signals through a correlation. 





\begin{figure}[t]
\sidecaption[t]
\includegraphics[width=7cm]{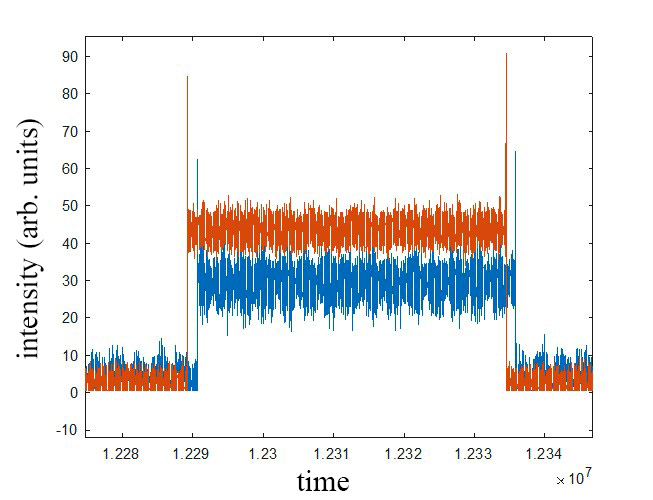}
\caption{Oscillator pulse, received by the two receptors at the same station. The delay between them is used to synchronized the signals. This pulse has a distinct duration, which makes it easy to distinguish from the transmitter's. From the delay, the position of the oscillator, and the orientation of the baseline of the 2-antennas station, the phase difference can be calculated.}
\label{fig:sync}       
\end{figure}

In order to generate this synchronization signal we employed a clock generator, Si5351A from Silicon Labs (Auston, USA) connected to a reference antenna. The oscillator generates pulses of a configurable frequency, with a cadence and duration that can also be adjusted. In particular, we set it to emit pulses of the same frequency as that of the transmitter, with a cadence of 10 seconds and a duration of 200~ms. The oscillator pulses, as well as the pulses from the transmitter carried by the animal, would be acquired by the processing software with some delay between receivers of each station due to the factors aforementioned. It is precisely this delay that we must correct to synchronize the signals with the purpose of deriving the phase difference. In this sense, then, the oscillator pulses during the measurements serve to synchronize the signals of each capture. Figure \ref{fig:sync} shows an oscillator pulse acquired by each receiver (one in red and one in blue) from the same receiving station. The pulse is the same but there is a delay between the receivers. 

Regarding the receivers' phase, RTL-SDR receivers feature a Phase-Locked Loop (PLL) with which, from a reference clock, they obtain the frequency to mix with the input signal. As a reference signal we used a 28.8~MHz clock, shared between the two receivers as mentioned above. The signals generated by each PLL, although they are referred to the same signal and therefore have the same frequency, do not necessarily have the same phase. Furthermore, every time the receivers are reset, the phase difference may change. This means that, although the signal received by the antennas travel the same path, the signals at the output of the receivers are out of phase and, furthermore, in each repetition of the experiment this phase would change. That is, there is a phase offset that we must correct. To achieve this, it is necessary to know the phase difference shown by a signal of the same frequency as that of the transmitter from a known location, for which we also used the reference oscillator. The fact that the frequency has to be the same is not obvious, and it follows from experiments in which we observed that the measured phase difference depends on the frequency of the emitted signal. So, we placed the oscillator at a known location and transmitted pulses of $150$~MHz. Calculating the phase difference of these pulses, we can know the value of the phase difference if the transmitter were placed at the site of the oscillator. This allows to calculate the necessary offset we need to apply to the transmitter pulses in order to correct the phase difference due to the PLL.

During the measurements in the forest we placed the oscillator at site C3, simulating pulses of the same frequency as that of the transmitter (with another duration and another cadence to identify them). To configure the frequency of the oscillator, we used the SDR\# software to check the pulses before each measurement and modified, if necessary, the frequency of the oscillator until it matched that of the transmitter (until they overlap in the spectrogram). This had to be done every night, since the frequency may be affected by external conditions (temperature, for example) and may not affect the transmitter and the oscillator in the same way.


\subsection{Phase difference estimation}

Once the captures of both receivers are synchronized, we can estimate the phase difference of the received signal. For this we take into account the following model. Let $S_1(t)$ and $S_2(t)$ be the signals received by antennas 1 and 2 (of the same pair), respectively:
\begin{align}
S_1(t) &= e^{i\omega t}, \\
S_2(t) &= e^{i\omega t+i\phi},
\end{align}
where $\phi$ is the phase difference between the signals, taking $S_1(t)$ as reference. Conjugating one of them and multiplying, we have:
\begin{align}
S^*_1(t)S_2(t) = e^{-i\omega t}e^{i\omega t+i\phi} = e^{i\phi},
\end{align}
from which the value of $\phi$ can be obtained.
From Eq.~(\ref{phase}), with $d=\lambda/2$, we can derive the direction of arrival, $\theta$, as:
\begin{align}
\theta = \arcsin\frac{\pi}{\phi}.
\label{eq:phasedif}
\end{align}

\subsection{Processing algorithm}

To acquire the signals, we started the capture of the receivers through two TCP interfaces set up in the capturing software. We usually captured during one minute, within which it is expected to receive a pulse from the transmitter every 4 seconds and one of the oscillator every 10 seconds. As explained above, the signals were aligned using the delay that there is, for the same pulse of the oscillator, between one receiver and the other of the same recording station. Once aligned, we filtered them with a low pass filter centered on the transmitter frequency, to reduce the noise and identify the pulses. The detected pulses were cropped from the entire acquisition and each one was subjected to a narrower filter (order 50) to further reduce signal noise. We used these signals, measured by each receiver, to calculate the product $S^*_1(t)S_2(t)$ and obtain the phase difference between receivers, of the received pulse. Let us remember that the value obtained in this way has to be further corrected using the reference pulse that the oscillator emits from a known location.

Using the phase difference information to obtain the direction of arrival and knowing the locations and baseline azimuths of the three pairs of antennas it is possible to estimate the point from which the pulse was transmitted by performing a triangulation (intersection of the arrival straight lines).

\section{Results and final considerations}
\label{sec:results}
\begin{figure}[t]
\includegraphics[width=\textwidth]{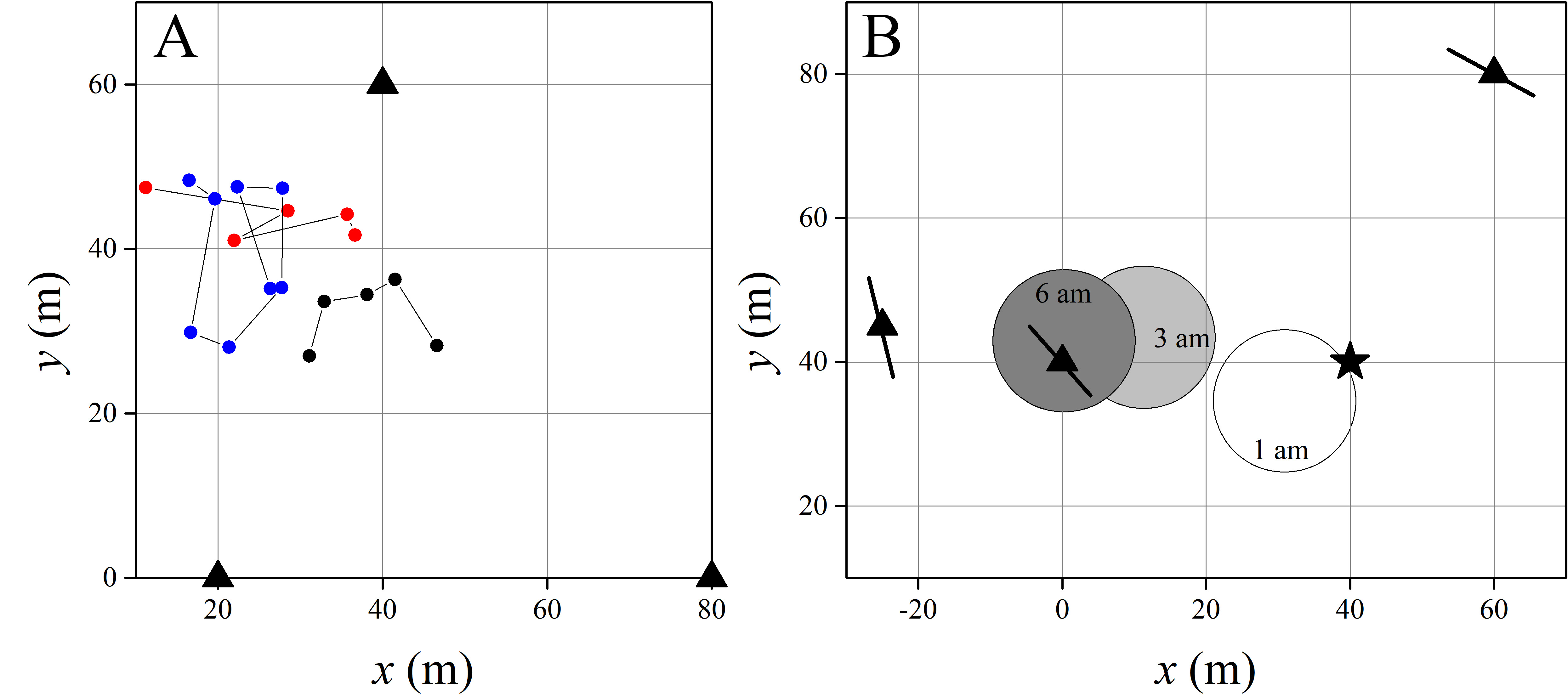}
\caption{A: Trajectories as measured on three consecutive nights (color circles) with the trilateration method. The large triangles show the position of the receiving stations. B: Regions visited during three periods of the night, as measured by the phase-difference method. Triangles show the position of the receiving stations (with a line showing the baseline), and a star shows the position of the oscillator.}
\label{fig:trajectories}       
\end{figure}

In this chapter, we showed two radiotelemetry techniques that were used to monitor animal movement in their natural habitat, primarily in hard-to-reach areas with dense vegetation. 
The trilateration method, applied to all captures, determines corresponding trajectories. In this section we show part of the results corresponding to the monitoring of an individual during three nights. Figure \ref{fig:trajectories} (A) shows part of the forest grid, the location of the three receiving stations (triangles) and trajectories made by an animal during 30 minutes of measurement, on three consecutive nights. Each point of the trajectories is the result of considering the maximum power values of each station given time windows of 60 seconds. It should be taken into account that it does not necessarily result in one point per minute, since there may be intervals in which we don't have any successful pulse. Using data dispersion during calibrations and preliminary measurements, we estimate the error of each point in 7 meters of radius. 

These results suggest that, with this methodology, it is possible to determine the regions through which the animal moves and reconstruct its trajectory with an error less than that obtained by working with GPS in the forest. This methodology would allow, for example, to determine the home range of the animal if we monitor its movement for several nights and for longer intervals. It would also be possible to determine hours of greatest and lowest activity, during the night and between nights, and to associate peaks of activity with external conditions (moonlight and temperature, among others). This methodology is also applicable to the monitoring of other species of small animals that live in the forest.

\begin{figure}[t]
\includegraphics[width=\textwidth]{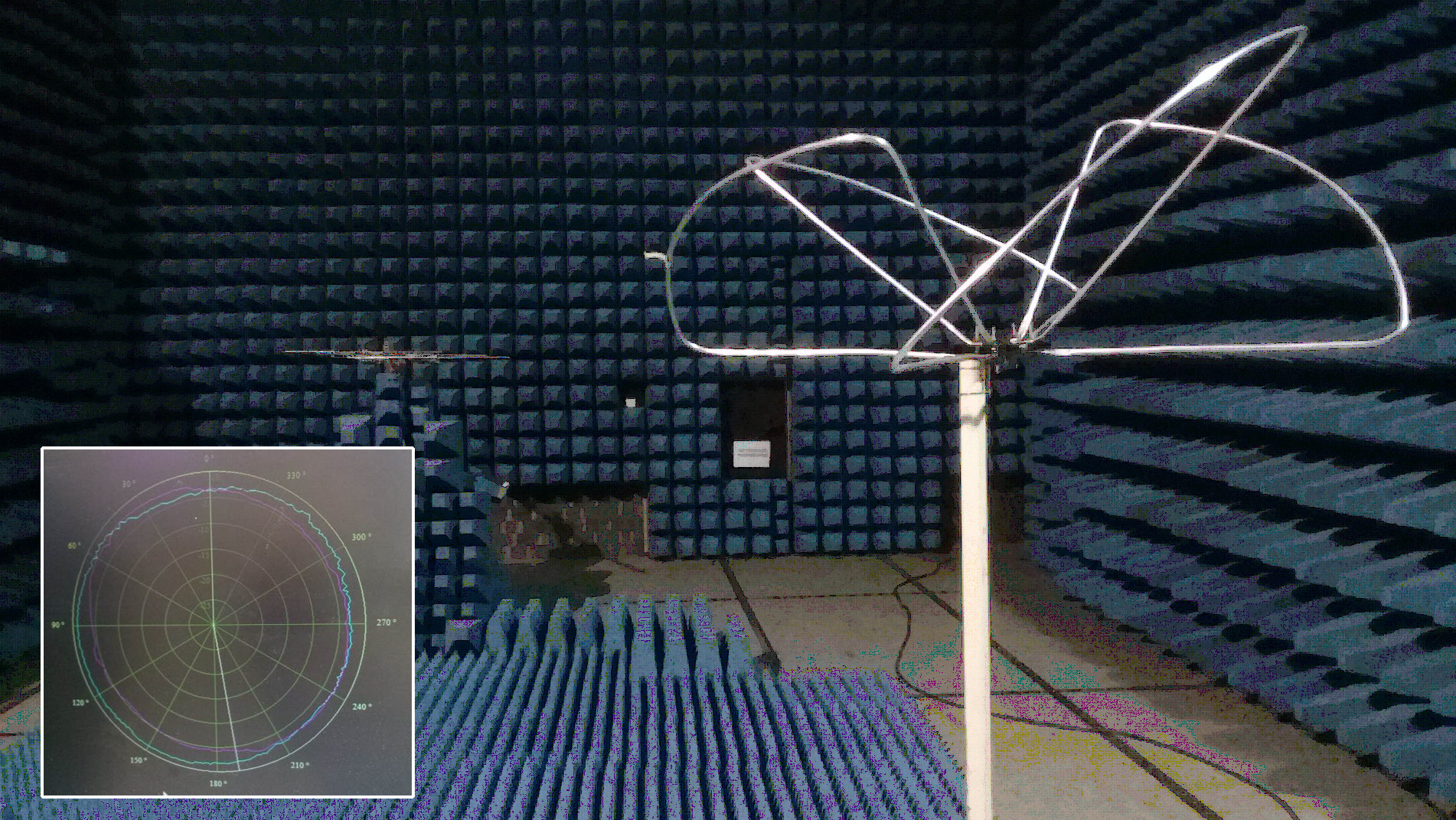}
\caption{Omnidirectional ``cloverleaf'' antenna, with circular polarization for the frequency 150 MHz, being tested in an anechoic chamber. Inset: radiation pattern in the horizontal plane, for horizontal polarization (purple) and vertical polarization (cyan).}
\label{fig:cloverleaf}       
\end{figure}

One way to improve the power measurement method is to use a different type of antenna. Since the omnidirectional antennas that we used are linearly polarized, they are sensitive to changes in the orientation of the transmitting antenna, as the animal moves in the forest. This is a major limitation of the methodology, since a change in polarization may be misinterpreted as a different distance of the transmitter from the antenna. This problem can be mitigated by using antennas with circular polarization, which would allow to decouple the changes in received power from possible changes in the relative orientation of the antennas. Since they are not commercially available, we designed, assembled and characterized our own omnidirectional circularly polarized antennas. They have a cloverleaf shape and were designed with the CST simulation software \cite{cst} (see Fig.~\ref{fig:cloverleaf}). 
The pattern shown in the inset corresponds to the radiation in the horizontal plane of a scaled-down prototype designed to operate at $915$~MHz, for both vertical (blue) and horizontal (purple) polarizations. We can see an axial ratio smaller  than $5$~dB. 

\begin{figure}[t]
\sidecaption[t]
\includegraphics[width=7cm]{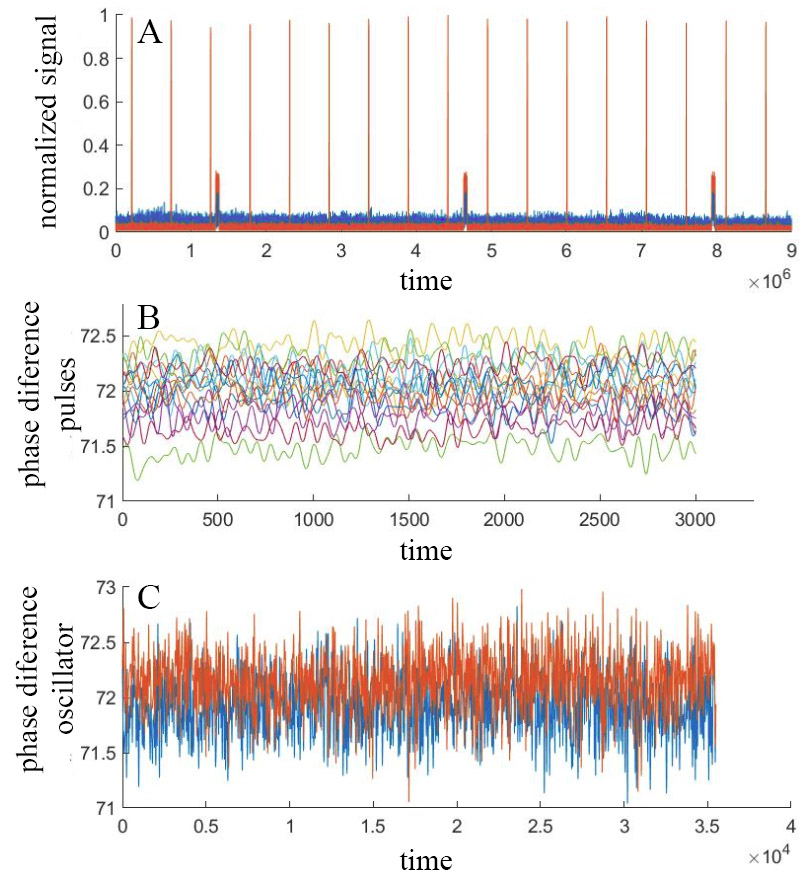}
\caption{A: Raw signal with pulses from the transmitter, including also 3 pulses from the oscillator, easily distinguished. Phase difference for each pulse from the transmitter (B) and for two pulses from the oscillator (C). The phase differences coincide at $72^\circ$ because they were transmitted from the same place.}
\label{fig:phasedif}       
\end{figure}

Regarding the phase difference measurement, Figure \ref{fig:phasedif} (A) shows an example of a capture taken by one of the monitoring stations during a test run in the forest. The most intense periodic pulses that repeat more frequently (one pulse every 2 seconds) correspond to the  transmitter. There are also three pulses that correspond to the reference oscillator pulses. Red color shows one of the receivers and blue the other one. The calculated phase difference of the transmitter and oscillator pulses is shown in Figure \ref{fig:phasedif} (B and C). In particular, if the oscillator and transmitter were located at the same point, the phase differences should be equal. Indeed, we can see that both values remain around $72^\circ$. It is because of this similarity that we can use the oscillator pulses as a reference to correct the offset. We performed a test, scanning angles of arrival from $0^\circ$ to $180^\circ$ every $20^\circ$, and checked that the relation between phase difference and angle of arrival is the expected sinusoidal sinusoidal of Eq.~(\ref{eq:phasedif}). The error remains around $\pm10^\circ$, which is an acceptable value to determine the direction of arrival.

We monitored the movement of the animals during the nights and also some days. The measurement periods varied between 4 hours (from 8pm to midnight) and 10 hours (from 8pm until 7am). Figure \ref{fig:trajectories} (B) shows the result of monitoring an individual in the field throughout the night. The animal was released at the position marked with a star. The circles represent the regions visited during three time ranges

Each pair of records of a station contains the pulses received by each antenna, from which we extract the maximum power value in a given time interval for each one, and the value of phase difference by comparing the phases of the pulses between the pair of antennas. The power and phase difference methodologies are complementary and, therefore, we can reconstruct the trajectory of the animals monitored each night.


This project constitutes a first step in our efforts to evaluate the role of potential intraspecific interaction between individuals on their distribution and use of space, and to test hypotheses about the resource dynamics and other behaviors of \emph{D. gliroides}. Their strong dependence on densely vegetated habitats makes them particularly sensitive to disruptions, such as deforestation and other human activities. In this regard we believe that the key role they play makes them a very relevant species for the conservation of the Patagonian forest. Additional knowledge of this ecosystem will certainly strengthen the conservation efforts. We have also began to develop and deploy related radiotelemetry systems for the study of other small animals in Patagonia.

\section*{Acknowledgements}
This research was supported by Consejo Nacional de Investigaciones Científicas y Técnicas (CONICET, PIP 112-2022-0100160 CO) and Universidad Nacional de Cuyo (06/C045-T1). 

G. Abramson and L. D. Kazimierski would like to thank the Isaac Newton Institute for Mathematical Sciences, Cambridge, for support and hospitality during the programme Mathematics of movement: an interdisciplinary approach to mutual challenges in animal ecology and cell biology, where work on this paper was undertaken. This work was supported by EPSRC grant no EP/R014604/1.

\end{document}